# Independent circuits in basal ganglia and cortex for the processing of reward and precision feedback

David Pascucci[1, 2], Clayton Hickey[1], Jorge Jovicich[1] and Massimo Turatto[1]

[1]Center for Mind/Brain Sciences (CIMeC), University of Trento, Rovereto, Italy
[2]Department of Psychology, University of Fribourg, Fribourg, Switzerland

**Abstract**

In order to understand human decision making it is necessary to understand how the brain uses feedback to guide goal-directed behavior. The ventral striatum (VS) appears to be a key structure in this function, responding strongly to explicit reward feedback. However, recent results have also shown striatal activity following correct task performance even in the absence of feedback. This raises the possibility that, in addition to processing external feedback, the dopamine-centered “reward circuit” might regulate endogenous reinforcement signals, like those triggered by satisfaction in accurate task performance. Here we use functional magnetic resonance imaging (fMRI) to test this idea. Participants completed a simple task that garnered both reward feedback and feedback about the precision of performance. Importantly, the design was such that we could manipulate information about the precision of performance within different levels of reward magnitude. Using parametric modulation and functional connectivity analysis we identified brain regions sensitive to each of these signals. Our results show a double dissociation: frontal and posterior cingulate regions responded to explicit reward but were insensitive to task precision, whereas the dorsal striatum - and putamen in particular - was insensitive to reward but responded strongly to precision feedback in reward-present trials. Both types of feedback activated the VS, and sensitivity in this structure to precision feedback was predicted by personality traits related to approach behavior and reward responsiveness. Our findings shed new light on the role of specific brain regions in integrating different sources of feedback to guide goal-directed behavior.

Corresponding author:

*David Pascucci*
Rue de Faucigny 2, 1700 Fribourg, Switzerland
Email: *david.pascucci@unifr.ch*
Phone: +41 (0)26 300 7627

## Introduction

Humans and other animals must be able to evaluate actions as a function of the quality of their outcome. Decades of neurophysiological and neuroimaging studies have demonstrated that the meso-cortico-striatal pathway is central to this function (McClure et al., 2004; O'Doherty, 2004; Schultz, 2000, 2006, 2013). Neurons in this system respond to explicit reward (Apicella et al., 1991; Knutson et al., 2003), signal errors in the prediction of reward (Schultz et al., 1997; Bayer & Glimcher, 2005), and drive selection of reward cues and approach toward these objects (Berridge & Robinson, 1998; Flagel et al., 2011; Hickey & Peelen, 2015). The ventral striatum (VS), a target of midbrain and cortical projections, has received particular attention in this context. This structure plays a core role in instrumental learning (O'Doherty et al., 2004) and reward-contingent behavior (Tricomi, Delgado & Fiez, 2004) and is sensitive to various types of external reward feedback (Knutson & Cooper, 2005).

The well-known sensitivity of the VS to reward feedback has led to the widely-held notion that this structure is in fact dedicated to the processing of reward. However, recent functional magnetic resonance (fMRI) findings have shown that the VS, together with other reward-related structures, is also activated by simple cognitive feedback such as that indicating performance accuracy (Rodriguez et al., 2006; Daniel & Pollmann, 2010; Tricomi & Fiez, 2008; Ullsperger & von Cramon, 2003; Han et al., 2010; Wolf et al. 2011).

Feedback-related responses in the striatum have been observed in a variety of tasks, ranging from information-integration learning (Daniel & Pollmann, 2010) to perceptual training (Tricomi et al., 2006). A handful of studies have observed striatal activation following accurate responses even when no explicit feedback is provided at all (Daniel & Pollmann, 2012; Satterthwaite et al. 2012; Guggenmos, Wilbertz, Hebart & Sterzer, 2016). In this situation, the VS responds most strongly when participants are completing a challenging task (Satterthwaite et al. 2012; Dobryakova, Jessup & Tricomi, 2016) or when they are confident about their performance (Daniel & Pollmann, 2012).

In addition to the VS, other striatal and cortical structures have been associated with both reward and performance processing. On one hand, the putamen - a key node in the motor feedback loop - responds to aspects of task performance that extend beyond purely motor execution processes. A number of studies have shown putamen activation in response to performance feedback (Cincotta & Seger, 2007; Eppinger, Schuck, Nystrom & Cohen, 2013), reward prediction errors (Garrison, Erdeniz & Done, 2013; Daniel & Pollmann, 2012; Sommer & Pollmann, 2016), performance evaluation and perceived competence, even in the absence of external feedback or reward (Daniel & Pollmann, 2010, 2012; Guggenmos et al., 2016; Sommer & Pollmann, 2016). On the other hand, regions such as orbitofrontal cortex (OFC) and posterior cingulate cortex (PCC) have been extensively linked to the processing of external reward (Liu, Hairston, Schrier & Fan, 2011). This suggests that performance feedback and internal signals of precision may target specific subcomponents of the reward system and striatal nuclei in particular. Reward-associated cortical areas, in contrast, may be sensitive to explicit primary and secondary reward feedback.

A number of studies have addressed the possibility that the dopaminergic system, and the striatum in particular, may contribute not only to the analysis of external rewards but also to the processing of internally-generated signals reflecting valuation of accurate performance (Satterthwaite et al. 2012; Daniel & Pollmann, 2012; Pascucci & Turatto, 2013; Pascucci, Mastropasqua & Turatto, 2015; see Daniel & Pollmann, 2014 for a review). For example, Daniel and Pollmann (2010) directly compared neural correlates of monetary reward with cognitive feedback during two parallel category-learning tasks. The authors found that both types of reinforcer activate the dopaminergic system in similar ways, but that a core structure of the VS, the nucleus accumbens (NAc), responded more strongly when learning was paired with monetary reward. Similarly, Delgado, Stenger and Fiez (2004) found that VS activation in response to the outcome of a gambling task was greater after reward-related feedback than after accuracy feedback, and Murayama et al. (2010) showed that the removal of external reward from a previously enjoyable task decreased the sensitivity of reward-related structures to task performance.

Taken together, this evidence suggests that reward incentives may be crucial in driving dopaminergic responses to performance outcomes. Tricomi and colleagues (Tricomi et al., 2006) have proposed that non-reward incentives like performance feedback become effective only under specific circumstances. As a result, motivational context and individual variability become important in predicting striatal sensitivity to different types of feedback (Tricomi et al., 2006; Delgado, Stenger & Fiez, 2004).

There is thus ambiguity in our understanding of striatal sensitivity to reward or performance feedback. One reason for this ambiguity is that existing studies investigating the role of non-reward information in striatal activation have understandably tended either to omit reward from the experimental design (Rodriguez et al 2006; Murayama et al., 2010; Daniel & Pollman, 2012; Satterthwaite et al. 2012) or have associated explicit reward to one task and accuracy feedback to another (Daniel & Pollmann, 2010; Delgado, Stenger & Fiez, 2004). Under these circumstances, it is unclear whether observed striatal sensitivity to task accuracy reflects a fundamental function of the area. It may be that this system always analyzes the quality of task performance, even when this kind of evaluation is not required by task instructions and is not required to achieve rewarding outcome. But it may alternatively be the case that, in the absence of external feedback, the dopaminergic system becomes sensitive to the next best learning signal, namely task accuracy.

Here we test these contrasting hypotheses. While in the fMRI scanner, we had human participants perform a simple video game that involved firing a bullet at a target. Each trial of this game resulted in one of five outcomes: a *perfect hit*, when the bullet hit the center of the target; a *good hit,* when the bullet hit the side of the target; a *near miss*, when the bullet hit the extreme edge of the target; a *near hit,* when the bullet just missed the target; and a *bad miss,* when the bullet landed far from the target (see Figure 1.B). Participants knew that hits resulted in monetary reward, but, critically, they were unaware that the game was rigged: the outcome of each trial was determined prior to task execution. This provided us the ability not only to manipulate whether a trial resulted in a hit, and thus whether reward was received, but also to vary the quality of the hit, and therefore the perceived precision of performance.

We used parametric analyses of the resulting fMRI data to isolate activity caused by the manipulation of explicit reward from activity caused by manipulation of task precision, and we used functional connectivity analysis to identify segregated networks supporting the processing of explicit reward feedback and task precision.

## Materials and Methods

### Subjects

Twenty healthy volunteers (mean age = 24 ± 3, 14 female) were recruited from the University of Trento and paid at the end of the experiment. All participants gave written informed consent. The study was conducted under the approval of the local institutional ethics committee.

**Visual stimulation**

Stimuli were back-projected onto a screen by a liquid-crystal projector at a frame rate of 60 Hz and a screen resolution of 1,280 × 1,024 pixels (mean luminance: 109 cd/m$^2$). Participants viewed the stimuli binocularly through a mirror above the head coil. Stimuli were generated with Matlab (Mathworks Inc., Natik, MA) and the Psychophysics Toolbox 3.8 (Pelli, 1997).

**Behavioral task**

Participants had to shoot a bullet (a red oval shape, 0.4° of diameter) from the top of a pointer (a small black rectangle, 2 x 0.5°) presented in the lower part of the display (at 10° from the center) by pressing a button on a response box. The target was a central white region (2.3 x 1°) of a black rectangle (7 x 1°) presented in the upper part of the display (at 8.5° from the center). Importantly, the bullet and pointer were horizontally jittered until the shot was fired (± 4° from the monitor's midline). The direction and speed of this movement jitter was varied randomly and the pointer and bullet were constantly sliding.

When the bullet was shot, it disappeared behind an occluder for a portion of its trajectory (gray rectangle, 10 x 16°). Behind this object, the bullet's trajectory was artificially deviated such that it reappeared in a trajectory that would land at a pre-determined position. We selected a set of five possible ending positions relative to the distance from the target's center (0±0.15° = perfect hit; 0.5±0.05° = good hit; 1±0.05° = near miss; 1.5±0.05° = near hit; 2.5±0.30° = bad miss; see Figure 1.B). Perfect hits, good hits, and near misses garnered 10 cents, with the other outcomes resulting in no gain (0 cents). The task thus defined 5 levels of task precision and two levels of reward feedback. Before the experiment, participants were instructed to focus on the position and speed of the pointer in order to select the right moment to shoot and were made explicitly aware that their performance dictated their earnings at the end of the experiment. At the end of the experiment, none of the participants reported being aware of the pre-determined nature of the game.

Each trial lasted 6.6 seconds and started with 800 ms of a green fixation spot (.5°) followed by the appearance of the task-related elements (see Figure 1.A). After one second, the bullet turned to light red and the gray central rectangle started moving. Following a change in the bullet's color, participants had one second to make a shot. After key press, the bullet moved toward the target and reached its final position in 1500 ms. When the bullet reached the target, the three elements became stationary and the outcome of the shot was shown for the rest of the trial. There were 50 trials in each run for a total of 250 trials. When participants failed to press the response button in time, the bullet fell from the pointer and the trial was discarded from analysis (less than 10% of trials discarded in total). The experimental session lasted approximately 30 minutes. Before the experiment, all participants underwent a brief practice session (20 trials) outside the scanner. During this practice session, each trial was followed by visual feedback indicating the reward obtained ("+10 cents" or "0 cents"). No feedback was provided inside the scanner.

**fMRI data acquisition**

FMRI images were acquired using a 4T Bruker MedSpec Biospin MR scanner and an 8-channel birdcage head coil. Each functional run consisted of 154 volumes with 32 T2*weighted echo planar slices (EPIs; repetition time (TR) = 2200 ms; time to echo (TE) = 30 ms; flip angle (FA) = 76°; field of view (FOV) = 192 x 192 $mm^2$; voxel size = 3 x 3 x 3 $mm^3$). EPI images were corrected for geometric distortions using the point-spread function method (Zaitsev et al., 2004). Before the experimental session, for each participant we acquired a structural whole-head image (MP-RAGE; TR = 2700 ms; TE = 4.18 ms; FA = 7°; FOV = 256 x 224 $mm^2$; inversion time (TI) = 1020 ms; voxel size = 1 x 1 x 1 $mm^3$; sagittal slices = 176) that was used for co-registration with the functional images.

**fMRI data preprocessing**

Anatomical and functional images were preprocessed with the Statistical Parametric Mapping toolbox (SPM12; University College of London, London, United Kingdom). The first four volumes of each run were discarded to allow for T1-equilibration effects. Functional images were then corrected for acquisition delay using the physical midpoint of acquisition as a reference. To correct for motion all images were realigned to the mean functional image using a two-pass procedure. Six motion parameters were obtained from the realignment procedure and were included in general linear model (GLM) analysis, which is described below. The anatomical scan was then co-registered to the mean image of the realigned functional volumes. Anatomical and functional images were subsequently normalized relative to the standard Montréal Neurological Institute (MNI) space using trilinear interpolation and smoothed with an isotropic 8 $mm^2$ full-width half-maximum Gaussian kernel.

As a final step, outlier volumes for each run (less than 5% on average) were identified through the compound-movement index available in the ART Toolbox (www.nitrc.org/projects/artifact_detect/; threshold = 2.5). A high-pass filter with cutoff of 128 seconds was applied to the time-series of functional images in order to remove low-frequency noise.

**Parametric modulation analysis**

Statistical analysis of functional images was performed using SPM12 and a set of custom scripts in Matlab. To investigate functional areas specialized in the processing of performance feedback (precision) and monetary reward, we used a model-based parametric modulation approach (Rohe, Weber & Fliessbach, 2012). Images were submitted to a two-stage mixed-effects model (Friston et al., 1994) with a single event of interest - the outcome of the shooting task - modeled with a delta function (duration = 0 seconds) and convolved with the canonical hemodynamic response function (HRF).

To model the variability in the strength of the neural response to the outcome as a function of the monetary win or the precision of performance, two parametric modulators were added to the event of interest. One modulator (Reward) was a stepwise function modulating the outcome regressor with a positive weight when a trial resulted in financial gain and a negative weight when it did not. The other modulator (Precision) was a discrete variable linearly increasing from bad misses to perfect hits in five steps (see Figure 1).

Because reward feedback occurred only when precision was high, the Reward and Precision factors were highly correlated. We accordingly constructed two separate general linear models (GLM) in which the two modulators were inverted and serially orthogonalized (Mumford, Poline & Poldrack, 2015). In the reward-first GLM, the Reward modulator was included before Precision, and, as a result, the Precision factor only explained variance not already explained by Reward. In the precision-first GLM this was reversed, such that the Reward factor only explained variance not already explained by Precision. This allowed us to disentangle functional areas responding uniquely to reward or precision feedback after accounting for variance shared by the two modulators. We subsequently examined this shared variance in the first factor of the two models (i.e. reward in the reward-first model and precision in the precision-first model) in order to identify areas sensitive to both types of feedback.

We performed an additional GLM to confirm the effect of Precision independent of Reward. Here, Reward was modeled as a two-level factor (reward vs. no reward). Within each level of this factor, the degrees of Precision were included as separate parametric modulators. This allowed us to examine the effect of precision feedback when reward was received (i.e. variance created by perfect hit, good hit, and near miss feedback when reward was received) and when it was not (i.e. variance created by near hit and bad miss feedback when no reward was received).

As a further step, to address whether activity in the putamen was driven exclusively by the precision feedback or by the interaction between the Reward and Precision factor (see Results), we adopted a model comparison approach (Rohe, Weber & Fliessbach, 2012). Two additional GLMs were constructed and estimated on the right posterior putamen seed, using an inclusive mask obtained from the significant cluster in the reward-first GLM.

Both models contained a single parametric modulator of the outcome. In the precision GLM, the modulator was the orthogonalized version of the Precision factor (from the reward-first GLM). In the reward-precision GLM, the modulator was an interaction term, obtained as the product of the Reward and Precision factor. We evaluated whether activity in the putamen was better explained by the Precision modulator or by its interaction with Reward by comparing the goodness of fit of the two models. The goodness of fit was estimated as the logarithm of the models residual variance ($log(\sigma^2)$), which represents a linear transformation of the Akaike information criterion (AIC) for models with equal number of data points and parameters (Rohe, Weber & Fliessbach, 2012). Individual $log(\sigma^2)$ values were then averaged across voxels of the putamen seed, separately for the precision and the reward-precision GLM. Model comparison was implemented at the group level by testing the difference between the $log(\sigma^2)$ of the models across participants (paired *t* test, two tails).

Six motion parameters derived from realignment were included in all GLMs as nuisance regressors, as were time and dispersion derivatives for each regressor, five constant terms defining the scanner runs, and a dummy variable coding for outlier volumes. The map of voxel-wise parameter estimates (beta values) for each regressor was obtained at the single-subject level. The beta images for the two orthogonalized modulators in precision-first GLM and reward-first GLM and for the two precision modulators in the parametric GLM were then submitted to second-level group analysis consisting of voxel-wise comparison across subjects (one-sample *t*-test), treating each subject as a random effect.

Statistical significance was assessed at the group-level using statistical non-parametric mapping (SnPM) which corrects for multiple comparisons at a *p*(FWE) < 0.05 (cluster-forming threshold = $p < 0.001$; number of permutations = 5000; no variance smoothing). To identify areas where neural activity was significantly explained by both Reward and Precision, conjunction analysis testing against the conjunction null hypothesis (*p*(FWE) < 0.05, Nichols et al., 2005) was performed using a second-level one-way ANOVA on individual statistical maps derived from the non-orthogonalized versions of the Reward and Precision modulators (the first parametric modulators of the reward-first and precision-first GLMs, respectively).

**Functional connectivity**

As a complement to the parametric modulation analysis, we investigated the functional connectivity at the time of the outcome for brain regions showing stronger sensitivity to reward or precision feedback. The goal of this additional analysis was to determine whether areas responding to reward or precision feedback were embedded in functionally segregated networks.

To this end, we used a generalized psychophysiological interaction approach (gPPI, McLaren, Ries, Xu & Johnson, 2012), which has the advantage over standard PPI procedures of accommodating multiple task conditions - including parametric modulators - in the same connectivity model (McLaren et al., 2012).

The aim of the gPPI analysis was to identify reward- and precision-related connectivity between seed regions of interest and the rest of the brain. One seed region (right posterior putamen) was defined as the cluster with the strongest effect of Precision in the reward-first GLM. Two other seeds (posterior cingulate, PCC; and medial orbitofrontal cortex, mOFC) were clusters with the strongest effect of Reward in the precision-first GLM. The first eigenvariate of the time-series of each seed was adjusted for the effects of interest and deconvolved from the HRF to estimate the time course of neuronal activity (Gitelman, Penny, Ashburner & Friston, 2003). Estimated neuronal time-series were then used to generate psychophysiological interactions with the main regressors of the reward-first and precision-first GLMs. Our interaction terms of interest were 1) the product of the right putamen neuronal time-series and the orthogonalized version of the precision modulator (from reward-first GLM) and 2) the product of the PCC and mOFC time-series and the orthogonalized version of the reward modulator (from precision-first GLM).

Psychophysiological interactions were reconvolved with the HRF and entered into three new GLMs. The putamen-precision GLM contained all regressors from the reward-first GLM along with the interaction terms with the putamen seed and its original eigenvariate time-series. The mOFC-reward PCC-reward GLMs contained all regressors from the precision-first GLM along with the interaction terms for the two reward seeds and their eigenvariate time-series. The inclusion of all regressors plus the seed eigenvariates allowed us to identify whole-brain connectivity driven by the effect of reward or precision modulators on the seed of interest, all while taking into account the main effect of the modulator and of the seed activity alone.

For each subject, three contrasts were computed from the gPPI models. In one contrast, we extracted beta values for the interaction between putamen activity and Precision (from the putamen-precision GLM). In the other two contrasts, we extracted beta values for the interactions between mOFC and Reward (from the mOFC-reward GLM) and between PCC and Reward (from the PCC-reward GLM). In line with our univariate approach, the construction of psychophysiological interactions with the orthogonalized version of each modulator allowed us to identify areas where a modulatory contribution of the seed activity depended on the unique effect of Precision or Reward feedback.

Individual contrasts were then submitted to second-level group analysis. Because only unshared variance among the three regressors (the orthogonalized Precision/Reward feedback, the seed activity and their interaction) loaded on the interaction term, the PPI analysis has implicitly less power than canonical univariate approaches and therefore, we assessed statistical significance with a whole-brain uncorrected threshold of $p = 0.001$ and cluster size of six voxels or greater.

**Questionnaire**

In a post-experimental session, eighteen participants were administered the Italian version (Leone, Pierro & Mannetti, 2002) of the behavioral inhibition (BIS) and behavioral approach (BAS) personality scale (Carver & White, 1994). The questionnaire consists of 24 Likert-scale questions (4 of which are fillers) assessing BIS (7 items) and three BAS subscales (Drive, Reward Responsiveness and Fun Seeking, 13 items). The BIS scale measures reaction to punishment, anxiety and response to stimuli inducing behavioral inhibition and withdrawal. The BAS scale measures reward responsiveness and reward seeking, representing individual differences in sensitivity to goal achievement, reward cues and approach behavior.

The BIS and BAS-total scores (the sum of scores in the three BAS subscales) were not correlated across participants (r = -0.21, $p$ = 0.39) and were used to predict inter-subject variability in response to performance feedback in reward-related regions. More precisely, we identified a region of interest (ROI) from the conjunction analysis and extracted beta values for the Precision modulator of the Reward Present trials in the parametric GLM (with the Reward factor held constant). Beta values were then fit to a linear model with standardized BIS and BAS-total scores as main predictors plus an intercept term.

## Results

### Precision

In the parametric modulation analysis of the reward-first GLM, the Precision modulator could account only for variance not already partitioned to the Reward manipulation. This revealed a single significant cluster of 66 voxels in the right posterior putamen (peak activity at x = 27, y = -4, z = -7, $T$ = 6.44; see Figure 2.A, green color scale, and Table 1). The sensitivity of this caudal portion of the striatum to precision feedback was corroborated by results from the parametric GLM. The analysis of the precision modulator in Reward Present trials identified two significant clusters located in the right posterior putamen (x = 27, y = -13, z = 2, $T$ = 5.23; see Figure 2.A, winter color scale, and Table 1) and left supramarginal gyrus (x = -57, y = -43, z = 20, $T$ = 5.70) where activity increased as a function of precision when reward was kept constant. No significant clusters were found for the precision modulator in Reward Absent trials.

To further characterize the pattern of results from the parametric GLM, we evaluated whether the interaction between precision and reward (i.e., the increasing effect of precision only under Reward Present trials) could represent a more reliable predictor of putamen activity than the Precision factor itself. The results of our model comparison revealed smaller $log(\sigma^2)$ values (see Methods) for the reward-precision GLM compared to the precision GLM ($T_{19}$ = 3.13, $p$ = 0.005), indicating that the interaction between reward and precision feedback provides a better model of putamen activity than the pure precision feedback (Figure 2.D).

The right posterior putamen cluster identified in the reward-first GLM was used to define the seed for analysis of functional connectivity. The results of the gPPI analysis revealed two separate clusters showing activity that correlated with the right putamen as a function of precision feedback. One cluster was located in the midbrain (peak of activity at x = 3, y = -19, z = -13, *T* = 4.65), including the ventral tegmental area (VTA) as identified in previous work (O'Doherty, Deichmann, Cricthley & Dolan, 2002; Bunzeck & Düzel, 2006; Krebs, Heipertz, Schuetze & Duzel, 2011). A second cluster was located in the supramarginal gyrus (x = 54, y = -40, z = 11, *T* = 3.94).

**Reward**

The precision-first GLM revealed two significant clusters where brain activity increased for monetary win, independent of precision feedback (see Figure 3.A, and Table 2). One cluster (84 voxels) was located in the medial part of the orbitofrontal cortex (mOFC; peak activation at x = 6, y = 65, z = -7, *T* = 4.99). The second cluster (121 voxels) included aspects of posterior cingulate cortex (PCC) and precuneus cortex (peak activation at x = 3, y = -55, z = 20, *T* = 5.74).

Two separate seeds were defined from these clusters and submitted to gPPI analysis (see Methods). No significant clusters were detected in these analyses.

**Reward and Precision**

To investigate regions where outcome-related activity covaried with both monetary reward and performance precision, we run a group conjunction analysis (see Materials and Methods) on statistical maps corresponding to the first factors in each of the reward-first and precision-first GLMs.

The conjunction analysis revealed two significant clusters in the left NAc (peak activation at x = -9, y = 8, z = -10, *T* = 6.78; see Figure 4.A, and Table 3) and right NAc (x = 15, y = 5, z = -10, *T* = 6.04), along with one cluster in the PCC (x = 3, y = -37, z = 29, *T* = 7.08) and one in the subcallosal cortex (x = 0, y = 14, z = -1, *T* = 6.94). NAc activity was thus elicited by both monetary and precision feedback.

To investigate whether this conjoined activation could underlie inter-subject variability in the responsiveness to precision, a linear regression model was used to predict NAc beta values for the Precision modulator in Reward Present trials of the parametric GLM based on individual measures of BIS and BAS-total. A significant regression model ($F(1,15) = 5.33$, $p = 0.017$, adjusted $R^2 = 0.338$) showed a non-significant intercept ($\beta = 0.075$, $p = 0.56$) and no predictive role for BIS ($\beta = -0.002$, $p = 0.98$), but a significant predictive role for BAS-total ($\beta = 0.426$, $p = 0.006$; see Figure 4.B). When reward was received, NAc was thus more sensitive to precision feedback in participants with high BAS-total scores.

## Discussion

We investigated brain areas involved in the processing of reward and performance feedback when both signals were present in the same task. To date, effects of accuracy feedback on striatal activity have been investigated in two ways: 1) with external reward explicitly omitted from an experimental design (Rodriguez et al 2006; Murayama et al., 2010; Daniel & Pollman, 2012; Satterthwaite et al. 2012), and 2) with reward and accuracy feedback alternated in separate blocks of trials (Daniel & Pollmann, 2010; Delgado, Stenger & Fiez, 2004). Results from both designs show that reward-related structures, and the VS in particular, respond to task accuracy and performance feedback. However, because these designs provide reward feedback or performance feedback exclusively, and never both at the same time, it is unclear whether the response to performance feedback in reward-related regions reflects a core function of these regions, or a secondary property that emerges only in the absence of external reward.

In an attempt to test the latter possibility, we had participants complete a video game designed such that both the magnitude of reward feedback and the perceived quality of task performance could be manipulated. Our analysis revealed two main findings. First, we found a double dissociation between sensitivity to reward and precision in the mOFC/PCC and dorsal striatum. This suggests specialized circuits for the processing of precision vs. monetary reward feedback. Second, we observed that the VS was sensitive to both precision and monetary feedback, and that the degree of VS sensitivity to precision feedback correlated with personality traits tied to motivation and reward responsiveness.

The mOFC sensitivity to reward is consistent with the established role of medial and central orbitofrontal regions in encoding the reward value of stimuli (Kim, Shimojo & O'Doherty, 2006; O'Doherty, 2007; Tsujimoto, Genovesio & Wise, 2011; Padoa-Schioppa & Assad, 2006). Furthermore, these structures are deeply involved in tracking monetary outcomes and receipt of reward (Knutson et al., 2001, 2003; Tremblay & Schultz, 2000; Rohe, Weber & Fliessbach, 2012). Using a parametric approach similar to the one presented here, Rohe and colleagues (2012) demonstrated that OFC preferentially signals whether or not a reward has been obtained, whereas prediction error and anticipatory signals emerge in striatal nuclei. In line with this finding, here we see that the medial portion of OFC responds uniquely to monetary gains, irrespective of precision feedback.

Reward outcome also modulated activity in regions of the PCC. Though the primary function of PCC remains unclear (Pearson et al., 2011), it appears to play a role in signaling behaviorally relevant events (Hayden, Smith & Platt, 2009; McCoy et al., 2003) such as the occurrence of reward in learning contexts (Hayden, Nair, McCoy & Platt, 2008; Leech & Sharp, 2014; Pearson et al., 2011). Recent work suggests that a key function of PCC may be to track and integrate the history of reward and behavior, promoting changes when actions do not lead to reward (Pearson, Hayden, Raghavachari & Platt, 2009). This is consistent with our results, where PCC activity may reflect a continuous process of action-outcome evaluation based on the reward gained on each trial.

Manipulation of precision feedback discretely activated the right dorsal putamen. This region, including the dorsocaudal sector of the striatum, has been defined as a key node within the cortical-basal ganglia motor loop (Ell, Hélie & Hutchinson, 2012). However, results also suggest that the role of the putamen extends into a wide range of cognitive functions, including working and episodic memory, cognitive control, category learning, habits learning and stimulus-response-outcome associations (see Ell and colleagues, 2012, for review). In particular, studies of both primates and humans converge to indicate that the putamen activation correlates with reward anticipation (McClure, Berns & Montague, 2003), reward magnitude (Cromwell & Schultz, 2003) and reward delivery (McClure et al., 2003).

In line with previous work (Cincotta & Seger, 2007; Eppinger, Schuck, Nystrom & Cohen, 2013; Daniel & Pollmann, 2010, 2012; Guggenmos et al., 2016; Sommer & Pollmann, 2016), our results support the idea that the putamen is involved in endogenous reward processing and performance valuation. This finding lends itself to two possible interpretations. One possibility is that sensitivity to precision in the posterior striatum reflects a genuine function of this structure that is independent of its well-known role in motor processing. This interpretation may be partly supported by the (uncorrected) results of our PPI analysis, which show that precision feedback modulates the functional connectivity between posterior putamen and VTA, an area primarily involved in the coordination of dopaminergic signals related to reward and motivation (Pignatelli & Bonci, 2005). Although putamen projections to the midbrain are mostly confined to the substantia nigra, studies in monkeys have shown that putamen nuclei receive input from VTA (Haber, Fudge & McFarland, 2000) and VTA-putamen functional connectivity has been reported in tasks with external reward (Krebs, Heipertz, Schuetze & Duzel, 2011). The posterior putamen may therefore be a key center for the analysis of endogenous reinforcers, relying on signals from the dopaminergic midbrain.

An alternative is that the involvement of putamen in performance monitoring and reward processing reflect the same underlying function. Performance-related signals in the putamen may indeed reflect feedback-driven updates of the strategy underlying a behavioral response (Monchi et al., 2006; Rubia et al., 2006; Ell et al., 2012). This possibility is in line with the well-established functional subdivision of the putamen during motor activity: the anterior part of the putamen is involved in the preparation of a movement whereas the posterior putamen is involved in execution (Gerardin et al., 2004; Jankowski et al., 2009). Thus, our results showing an effect of precision on posterior putamen may suggest an off-line process of reinforcement by which the neural pattern underlying the performed action is reinforced on the basis of precision feedback. Crucially, this motor reinforcement process in the putamen could be regulated by dopaminergic input from the VTA.

Recently, Tricomi and colleagues have shown that the right posterior putamen becomes more sensitive to the onset of task-related stimuli with increasing task experience (Tricomi, Balleine & O'Doherty, 2009). This result, combined with our findings, illustrates the critical role of the putamen in the development of stimulus-response associations and habitual behavior. Early on, the posterior putamen may process performance-related feedback, exploiting the consequences of every action. But later, once optimal stimulus-response associations are established, the putamen may shift its response to stimuli that anticipate the action, in order to select the learned response and promote reflexive and habitual behavior (Tricomi, Balleine & O'Doherty, 2009).

Although the right putamen was significantly modulated by precision feedback, we found that this modulation was mainly driven by precision in Reward Present trials, whereas no significant effect was found in Reward Absent trials. This interaction between reward and precision feedback in the putamen suggests that the hypothesized reinforcement mechanism only operates when performance feedback is provided within the context of correct, rewarded performance.

Our results demonstrate that the NAc is sensitive to both precision and reward feedback. A widely held view is that NAc activity, mediated by dopaminergic midbrain inputs, is sensitive to reward prediction-error signals, or discrepancies between expected and actual reward (Horovitz, 2009; Schultz, 2013; Knutson et al., 2001; O'Doherty et al., 2002; Fiorillo, Tobler & Schultz, 2003; Floresco, 2015). A growing literature indicates that similar prediction-error signals are generated in the NAc as a function of performance accuracy (Daniel & Pollmann, 2012; Satterthwaite et al. 2012; Guggenmos et al., 2016), reflecting endogenous valuation of task performance in the absence of external reward. A core goal of our experiment was to identify the primary sensitivity of the NAc – to find out which type of feedback it was most sensitive to. We considered two hypotheses: 1) that the VS activation is primarily driven by endogenous, performance-related signals, and 2) that the VS responds to endogenous reinforcers only when external reward is omitted from the task. We believed that combining monetary reward and performance feedback in the same paradigm would allow us to reveal the primary role of the NAc. However, our results show no reliable difference: in this task at least, NAc responds to precision and reward feedback in much the same way.

To further characterize this result, we tested the relationship between NAc sensitivity to precision feedback with results from a personality inventory. The possibility that NAc sensitivity to performance may be a product of personality has been discussed in the literature (e.g., Daniel & Pollmann, 2014), but never tested. Previous work provided an initial support to this idea by showing that perceived competence, as assessed through a motivation questionnaire, predicted NAc activation in response to cognitive feedback (Daniel & Pollmann, 2010). Our results provide novel evidence that personality traits play a role in mediating the sensitivity of NAc to precision feedback. Subjects with strong positive reactions to reward (high BAS-total scores) are sensitive to precision feedback, perhaps reflecting the treatment of this information as a type of endogenous reward signal.

We broadly interpret our results in terms of an actor-critic model of reinforcement learning. According to this, reinforcement learning requires a critic module that uses prediction-error signals to provide recurrent updates about the probability of external reward, and an actor module that encodes the causal link between stimuli, actions and reward, selecting the optimal behavior in order to gain reward in the future (Joel, Niv & Ruppin, 2002). Evidence from neurophysiology and neuroimaging shows that, through dopaminergic signaling, the VS (O'Dohery et al., 2004; Pessiglione et al., 2006) and the OFC (Kim, Shimojo & O'Doherty, 2006;

O'Doherty, Hampton & Kim, 2007) may act as the critic whereas the dorsal posterior striatum may represent the actor. Although the biological plausibility of the actor-critic model has been subject to criticism (Joel et al. 2002), such functional distinction is supported by a large body of literature linking specific regions of the cortico-striatal circuit, such as the anterior VS and the frontal cortex, to reward prediction and hedonic experience (Floresco, 2015), and the posterior putamen to motor-related processes, such as motor execution, planning and, in particular, motor learning (Joel et al., 2002; Tricomi et al., 2009). Our results support this functional subdivision by showing that the same visual information - a bullet hitting a target - can differentially trigger both reward-based reinforcement signals and performance-related modulation of ongoing brain activity.

In conclusion, by revealing its sensitivity to both endogenous and exogenous reinforcement signals, the present results support the idea that the VS plays a key role in feedback processing. At the same time, we report novel evidence of specificity in other structures – the putamen and mOFC in particular – where precision feedback and monetary reward are selectively processed. By working in concert, these regions appear to integrate information from varying feedback sources in order to guide future choices and to optimize behavior.

## References

Apicella, P., Ljungberg, T., Scarnati, E., & Schultz, W. (1991). Responses to reward in monkey dorsal and ventral striatum. Experimental Brain Research, 85(3), 491-500.

Bayer, H. M., & Glimcher, P. W. (2005). Midbrain dopamine neurons encode a quantitative reward prediction error signal. Neuron, 47(1), 129-141.

Berridge, K. C., & Robinson, T. E. (1998). What is the role of dopamine in reward: hedonic impact, reward learning, or incentive salience?. Brain Research Reviews, 28(3), 309-369.

Bunzeck, N., & Düzel, E. (2006). Absolute coding of stimulus novelty in the human substantia nigra/VTA. Neuron, 51(3), 369-379.

Carver, C. S., & White, T. L. (1994). Behavioral inhibition, behavioral activation, and affective responses to impending reward and punishment: The BIS/BAS Scales. Journal of personality and social psychology, 67(2), 319.

Cincotta, C. M., & Seger, C. A. (2007). Dissociation between striatal regions while learning to categorize via feedback and via observation. Journal of cognitive neuroscience, 19(2), 249-265.

Cromwell, H. C., & Schultz, W. (2003). Effects of expectations for different reward magnitudes on neuronal activity in primate striatum. Journal of Neurophysiology, 89(5), 2823-2838.

Daniel, R., & Pollmann, S. (2010). Comparing the neural basis of monetary reward and cognitive feedback during information-integration category learning. The Journal of Neuroscience, 30(1), 47-55.

Daniel, R., & Pollmann, S. (2012). Striatal activations signal prediction errors on confidence in the absence of external feedback. Neuroimage, 59(4), 3457-3467.

Daniel, R., & Pollmann, S. (2014). A universal role of the ventral striatum in reward-based learning: evidence from human studies. Neurobiology of learning and memory, 114, 90-100.

Delgado, M. R., Stenger, V. A., & Fiez, J. A. (2004). Motivation-dependent responses in the human caudate nucleus. Cerebral Cortex, 14(9), 1022-1030.

Dobryakova, E., Jessup, R. K., & Tricomi, E. (2016). Modulation of ventral striatal activity by cognitive effort. NeuroImage.

Ell, S. W., Hélie, S., & Hutchinson, S. (2012). Contributions of the putamen to cognitive function. In A. Costa & E. Villalba (eds.). Horizon in Neuroscience.

Volume 7 (pp. 29-52). Nova Publishers.

Eppinger, B., Schuck, N. W., Nystrom, L. E., & Cohen, J. D. (2013). Reduced striatal responses to reward prediction errors in older compared with younger adults. Journal of Neuroscience, 33(24), 9905-9912.

Fiorillo, C. D., Tobler, P. N., & Schultz, W. (2003). Discrete coding of reward probability and uncertainty by dopamine neurons. Science, 299(5614), 1898-1902.

Flagel, S. B., Clark, J. J., Robinson, T. E., Mayo, L., Czuj, A., Willuhn, I., ... & Akil, H. (2011). A selective role for dopamine in stimulus-reward learning. Nature, 469(7328), 53-57.

Floresco, S. B. (2015). The nucleus accumbens: an interface between cognition, emotion, and action. Annual review of psychology, 66, 25-52.

Friston, K. J., Holmes, A. P., Worsley, K. J., Poline, J. P., Frith, C. D., & Frackowiak, R. S. (1994). Statistical parametric maps in functional imaging: a general linear approach. Human brain mapping, 2(4), 189-210.

Garrison, J., Erdeniz, B., & Done, J. (2013). Prediction error in reinforcement learning: a meta-analysis of neuroimaging studies. Neuroscience & Biobehavioral Reviews, 37(7), 1297-1310.

Gerardin, E., Pochon, J. B., Poline, J. B., Tremblay, L., Van de Moortele, P. F., Levy, R., ... & Lehéricy, S. (2004). Distinct striatal regions support movement selection, preparation and execution. Neuroreport, 15(15), 2327-2331.

Gitelman, D. R., Penny, W. D., Ashburner, J., & Friston, K. J. (2003). Modeling regional and psychophysiologic interactions in fMRI: the importance of hemodynamic deconvolution. Neuroimage, 19(1), 200-207.

Guggenmos, M., Wilbertz, G., Hebart, M. N., & Sterzer, P. (2016). Mesolimbic confidence signals guide perceptual learning in the absence of external feedback. eLife, 5, e13388.

Haber, S. N., Fudge, J. L., & McFarland, N. R. (2000). Striatonigrostriatal pathways in primates form an ascending spiral from the shell to the dorsolateral striatum. Journal of Neuroscience, 20(6), 2369-2382.

Han, S., Huettel, S. A., Raposo, A., Adcock, R. A., & Dobbins, I. G. (2010). Functional significance of striatal responses during episodic decisions: recovery or goal attainment?. The Journal of Neuroscience, 30(13), 4767-4775.

Hayden, B. Y., Nair, A. C., McCoy, A. N., & Platt, M. L. (2008). Posterior cingulate

cortex mediates outcome-contingent allocation of behavior. Neuron, 60(1), 19-25.

Hayden, B. Y., Smith, D. V., & Platt, M. L. (2009). Electrophysiological correlates of default-mode processing in macaque posterior cingulate cortex. Proceedings of the National Academy of Sciences, 106(14), 5948-5953.

Hickey, C., & Peelen, M. V. (2015). Neural Mechanisms of Incentive Salience in Naturalistic Human Vision. Neuron, 85(3), 512-518.

Horvitz, J. C. (2009). Stimulus–response and response–outcome learning mechanisms in the striatum. Behavioural brain research, 199(1), 129-140.

Jankowski, J., Scheef, L., Hüppe, C., & Boecker, H. (2009). Distinct striatal regions for planning and executing novel and automated movement sequences.Neuroimage, 44(4), 1369-1379.

Joel, D., Niv, Y., & Ruppin, E. (2002). Actor–critic models of the basal ganglia: New anatomical and computational perspectives. Neural networks, 15(4), 535-547.

Kim, H., Shimojo, S., & O'Doherty, J. P. (2006). Is avoiding an aversive outcome rewarding? Neural substrates of avoidance learning in the human brain. PLoS Biol, 4(8), e233.

Knutson, B., & Cooper, J. C. (2005). Functional magnetic resonance imaging of reward prediction. Current opinion in neurology, 18(4), 411-417.

Knutson, B., Adams, C. M., Fong, G. W., & Hommer, D. (2001). Anticipation of increasing monetary reward selectively recruits nucleus accumbens. J Neurosci, 21(16), RC159.

Knutson, B., Fong, G. W., Bennett, S. M., Adams, C. M., & Hommer, D. (2003). A region of mesial prefrontal cortex tracks monetarily rewarding outcomes: characterization with rapid event-related fMRI. Neuroimage, 18(2), 263-272.

Krebs, R. M., Heipertz, D., Schuetze, H., & Duzel, E. (2011). Novelty increases the mesolimbic functional connectivity of the substantia nigra/ventral tegmental area (SN/VTA) during reward anticipation: evidence from high-resolution fMRI. Neuroimage, 58(2), 647-655.

Leech, R., & Sharp, D. J. (2014). The role of the posterior cingulate cortex in cognition and disease. Brain, 137(1), 12-32.

Leone, L., Pierro, A., & Mannetti, L. (2002). Validità della versione Italiana delle Scale BIS/BAS di Carver e White (1994): Generalizzabilità della struttura e relazioni con costrutti affini. Giornale Italiano di Psicologia, 29(2), 413-436.

Liu, X., Hairston, J., Schrier, M., & Fan, J. (2011). Common and distinct networks underlying reward valence and processing stages: a meta-analysis of functional neuroimaging studies. Neuroscience & Biobehavioral Reviews, 35(5), 1219-1236.

McClure, S. M., Berns, G. S., & Montague, P. R. (2003). Temporal prediction errors in a passive learning task activate human striatum. Neuron, 38(2), 339-346.

McClure, S. M., York, M. K., & Montague, P. R. (2004). The neural substrates of reward processing in humans: the modern role of FMRI. The Neuroscientist,10(3), 260-268.

McCoy, A. N., Crowley, J. C., Haghighian, G., Dean, H. L., & Platt, M. L. (2003). Saccade reward signals in posterior cingulate cortex. Neuron, 40(5), 1031-1040.

McLaren, D. G., Ries, M. L., Xu, G., & Johnson, S. C. (2012). A generalized form of context-dependent psychophysiological interactions (gPPI): a comparison to standard approaches. Neuroimage, 61(4), 1277-1286.

Monchi, O., Petrides, M., Strafella, A. P., Worsley, K. J., & Doyon, J. (2006). Functional role of the basal ganglia in the planning and execution of actions. Ann Neurol, 59(2), 257-264.

Mumford, J. A., Poline, J. B., & Poldrack, R. A. (2015). Orthogonalization of regressors in FMRI models. PloS one, 10(4), e0126255.

Murayama, K., Matsumoto, M., Izuma, K., & Matsumoto, K. (2010). Neural basis of the undermining effect of monetary reward on intrinsic motivation. Proceedings of the National Academy of Sciences, 107(49), 20911-20916.

Nichols, T., Brett, M., Andersson, J., Wager, T., & Poline, J. B. (2005). Valid conjunction inference with the minimum statistic. Neuroimage, 25(3), 653-660.

O'Doherty, J. P. (2004). Reward representations and reward-related learning in the human brain: insights from neuroimaging. Current opinion in neurobiology, 14(6), 769-776.

O'Doherty, J. P. (2007). Lights, camembert, action! The role of human orbitofrontal cortex in encoding stimuli, rewards, and choices. Annals of the New York Academy of Sciences, 1121(1), 254-272.

O'Doherty, J. P., Deichmann, R., Critchley, H. D., & Dolan, R. J. (2002). Neural responses during anticipation of a primary taste reward. Neuron, 33(5), 815-826.

O'Doherty, J. P., Hampton, A., & Kim, H. (2007). Model-based fMRI and its application to reward learning and decision making. Annals of the New York Academy of sciences, 1104(1), 35-53.

O'Doherty, J., Dayan, P., Schultz, J., Deichmann, R., Friston, K., & Dolan, R. J. (2004). Dissociable roles of ventral and dorsal striatum in instrumental conditioning. Science, 304(5669), 452-454.

Padoa-Schioppa, C., & Assad, J. A. (2006). Neurons in orbitofrontal cortex encode economic value. Nature, 441(7090), 223.

Pascucci, D., & Turatto, M. (2013). Immediate effect of internal reward on visual adaptation. Psychological science, 24(7), 1317-1322.

Pascucci, D., Mastropasqua, T., & Turatto, M. (2015). Monetary Reward Modulates Task-Irrelevant Perceptual Learning for Invisible Stimuli. PLoS ONE 10(5): e0124009.

Pearson, J. M., Hayden, B. Y., Raghavachari, S., & Platt, M. L. (2009). Neurons in posterior cingulate cortex signal exploratory decisions in a dynamic multioption choice task. Current biology, 19(18), 1532-1537.

Pearson, J. M., Heilbronner, S. R., Barack, D. L., Hayden, B. Y., & Platt, M. L. (2011). Posterior cingulate cortex: adapting behavior to a changing world. Trends in cognitive sciences, 15(4), 143-151.

Pelli DG (1997) The VideoToolbox software for visual psychophysics: transforming numbers into movies. Spatial Vision, 10, 437–442. pmid:9176953 doi: 10.1163/156856897x00366

Pessiglione, M., Seymour, B., Flandin, G., Dolan, R. J., & Frith, C. D. (2006). Dopamine-dependent prediction errors underpin reward-seeking behaviour in humans. Nature, 442(7106), 1042-1045.

Pignatelli, M., & Bonci, A. (2015). Role of dopamine neurons in reward and aversion: a synaptic plasticity perspective. Neuron, 86(5), 1145-1157.

Rodriguez, P. F., Aron, A. R., & Poldrack, R. A. (2006). Ventral–striatal/nucleus–accumbens sensitivity to prediction errors during classification learning. Human brain mapping, 27(4), 306-313.

Rohe, T., Weber, B., & Fliessbach, K. (2012). Dissociation of BOLD responses to reward prediction errors and reward receipt by a model comparison. European Journal of Neuroscience, 36(3), 2376-2382.

Rubia, K., Smith, A. B., Woolley, J., Nosarti, C., Heyman, I., Taylor, E., et al. (2006). Progressive increase of frontostriatal brain activation from childhood to adulthood during event-related tasks of cognitive control. Hum Brain Mapp, 27(12), 973-993.

Satterthwaite, T. D., Ruparel, K., Loughead, J., Elliott, M. A., Gerraty, R. T., Calkins, M. E., ... & Wolf, D. H. (2012). Being right is its own reward: Load and performance related ventral striatum activation to correct responses during a working memory task in youth. Neuroimage, 61(3), 723-729.

Schultz, W. (2000). Multiple reward signals in the brain. Nature reviews neuroscience, 1(3), 199-207.

Schultz, W. (2006). Behavioral theories and the neurophysiology of reward. Annu. Rev. Psychol., 57, 87-115.

Schultz, W. (2013). Updating dopamine reward signals. Current opinion in neurobiology, 23(2), 229-238.

Schultz, W., Dayan, P., & Montague, P. R. (1997). A neural substrate of prediction and reward. Science, 275(5306), 1593-1599.

Sommer, S., & Pollmann, S. (2016). Putamen Activation Represents an Intrinsic Positive Prediction Error Signal for Visual Search in Repeated Configurations. The Open Neuroimaging Journal, 10, 126.

Tremblay, L., & Schultz, W. (2000). Modifications of reward expectation-related neuronal activity during learning in primate orbitofrontal cortex. Journal of neurophysiology, 83(4), 1877-1885.

Tricomi, E. M., Delgado, M. R., & Fiez, J. A. (2004). Modulation of caudate activity by action contingency. Neuron, 41(2), 281-292.

Tricomi, E., & Fiez, J. A. (2008). Feedback signals in the caudate reflect goal achievement on a declarative memory task. Neuroimage, 41(3), 1154-1167.

Tricomi, E., Balleine, B. W., & O’Doherty, J. P. (2009). A specific role for posterior dorsolateral striatum in human habit learning. European Journal of Neuroscience, 29(11), 2225-2232.

Tricomi, E., Delgado, M. R., McCandliss, B. D., McClelland, J. L., & Fiez, J. A. (2006). Performance feedback drives caudate activation in a phonological learning task. Journal of cognitive neuroscience, 18(6), 1029-1043.

Tsujimoto, S., Genovesio, A., & Wise, S. P. (2011). Frontal pole cortex: encoding ends at the end of the endbrain. Trends in cognitive sciences, 15(4), 169-176.

Ullsperger, M., & Von Cramon, D. Y. (2003). Error monitoring using external feedback: specific roles of the habenular complex, the reward system, and the cingulate motor area revealed by functional magnetic resonance imaging. The Journal of neuroscience, 23(10), 4308-4314.

Wolf, D. H., Gerraty, R., Satterthwaite, T. D., Loughead, J., Campellone, T., Elliott, M. A. et al. (2011). Striatal intrinsic reinforcement signals during recognition memory: relationship to response bias and dysregulation in schizophrenia. Frontiers in behavioral neuroscience. 5(81), 10.3389/fnbeh.2011.00081.

Zaitsev, M., Hennig, J., & Speck, O. (2004). Point spread function mapping with parallel imaging techniques and high acceleration factors: Fast, robust, and flexible method for echo-planar imaging distortion correction. Magnetic Resonance in Medicine, 52(5), 1156-1166.

**Figure legends**

**Figure 1.** A) Example of events in a single trial of the task. During “aiming”, the bottom pointer and bullet were jittering and participants had 1000 ms to “shoot” the bullet and hit the white portion of the target (upper side of the screen). The outcome of each shot was shown until the end of the trial and represented the event of interest for the fMRI analyses. B) Outcomes (the ending positions of the bullet) were determined prior to task execution and divided into five levels of precision. Reward (10 cents) was delivered exclusively when the bullet hit the central white portion of the target, independent of precision. Reward and Precision were used as parametric modulators of brain activity at the time of the outcome. Stimuli are not drawn to scale.

**Figure 2.** A) Results of the parametric modulation analysis for the Precision modulator orthogonalized to Reward (reward-first GLM), shown in the green scale, and for the Precision modulator orthogonalized to monetary win (parametric GLM), overlaid in the winter color scale. B) Functional connectivity (gPPI) results for the right posterior putamen seed. Statistical maps are superimposed on a MNI ICMB152 Average Brain atlas using MRIcron software (www.mricro.com) and thresholded according to the corrections described in the Materials and Methods section. C) Average beta estimates in the right putamen peak from reward-first GLM as a function of Precision (reward-first GLM) and Reward (precision-first GLM). Bars are 95% confidence intervals for the mean. D) Mean logarithmic residual variance ($log(\sigma^2)$) of the GLMs estimated in the putamen seed, containing either the Precision modulator or the Reward x Precision interaction modulator. Smaller $log(\sigma^2)$ values represent superior model fit. The overall mean has been subtracted for graphical purpose. Bars are ±1 standard errors of the mean.

**Figure 3.** A) Results of the parametric modulation analysis for the Reward modulator orthogonalized to Precision (precision-first GLM). B) Average beta estimates in the PCC and mOFC maxima from precision-first GLM as a function of Precision (reward-first GLM) and Reward (precision-first GLM).

**Figure 4.** A) Results of the conjunction analysis with Reward (reward-first GLM) and Precision (precision-first GLM). B) Average beta estimates (blue triangles) of the effect of Precision on Reward Present trials (parametric GLM) in the NAc ROI as a function of BAS-total with the fitted regression line and 95 % confidence intervals for predicted responses.

**Table 1**

| Region Label | Extent | t-value | MNI Coordinates x | y | z |
|---|---|---|---|---|---|
| **Precision** (reward-first GLM) | | | | | |
| Right putamen | 66 | 6.444 | 27 | -4 | -7 |
| **Precision** (parametric GLM-Reward Present) | | | | | |
| Supramarginal Gyrus, posterior division | 162 | 5.707 | -57 | -43 | 20 |
| Angular Gyrus | 162 | 4.956 | -60 | -52 | 11 |
| Right putamen | 123 | 5.235 | 27 | -13 | 2 |
| **Right putamen/Precision gPPI Connectivity** | | | | | |
| Brain-Stem (VTA) | 6 | 4.649 | 3 | -19 | -13 |
| Supramarginal Gyrus, posterior division | 6 | 3.940 | 54 | -40 | 11 |

**Table 1.** Maxima of activation from the group statistic on the Precision modulator. Significant peaks are reported for the GLM with Precision orthogonalized with respect to Reward (reward-first GLM) and to monetary win (parametric GLM). Peaks of activity correlated with the interaction right posterior putamen x Precision (see Materials and Methods) are included at the bottom. The table includes the cluster size, coordinates and T-values of peaks separated by more than 10 mm. Only local maxima in uniquely-labeled grey matter regions are reported. Regions are labeled using the Harvard-Oxford maximum probability atlas (SPM).

**Table 2**

| Region Label | Extent | t-value | MNI Coordinates x | y | z |
|---|---|---|---|---|---|
| **Reward** (precision-first GLM) | | | | | |
| Precuneus Cortex | 121 | 5.746 | -3 | -55 | 20 |
| mOFC | 84 | 4.998 | 6 | 65 | -7 |
| Paracingulate Gyrus | 84 | 4.451 | 0 | 53 | -4 |

**Table 2.** Maxima of activation from the group statistic on the Reward modulator. Significant peaks are reported for the GLM with Reward orthogonalized with respect to Precision (precision-first GLM).

**Table 3**

| Region Label | Extent | t-value | MNI Coordinates x | y | z |
|---|---|---|---|---|---|
| **Reward & Precision** (conjunction analysis) | | | | | |
| Cingulate Gyrus, posterior division | 27 | 7.081 | 3 | -37 | 29 |
| Subcallosal Cortex | 8 | 6.945 | 0 | 14 | -1 |
| Left Accumbens | 34 | 6.787 | -9 | 8 | -10 |
| Right putamen | 5 | 6.048 | 15 | 5 | -10 |

**Table 3.** Maxima of activation from the group conjunction analysis of the non-orthogonalized version of the Reward (reward-first GLM) and Precision (precision-first GLM) modulators.

**Figure 1**

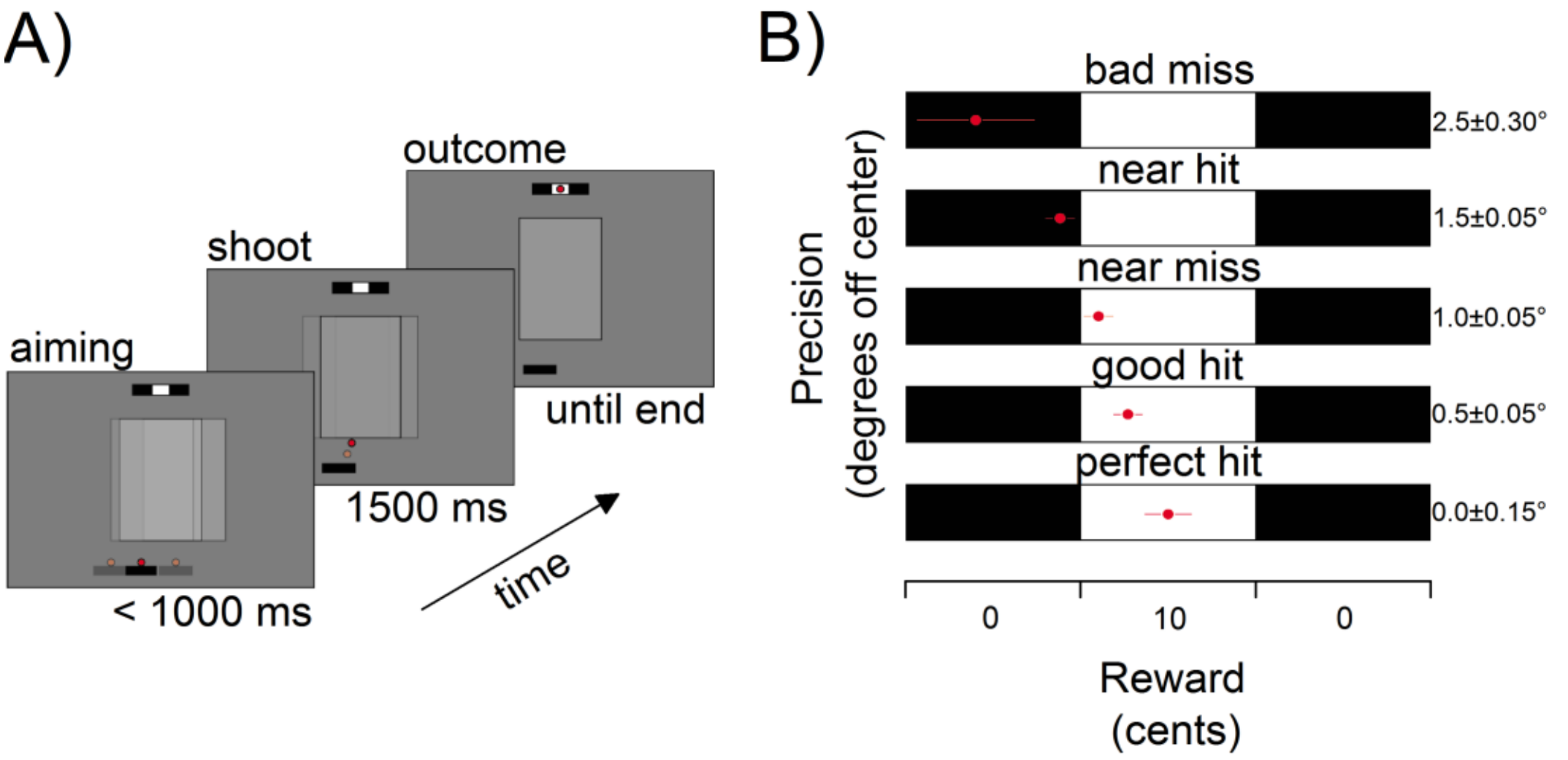

**Figure 2**

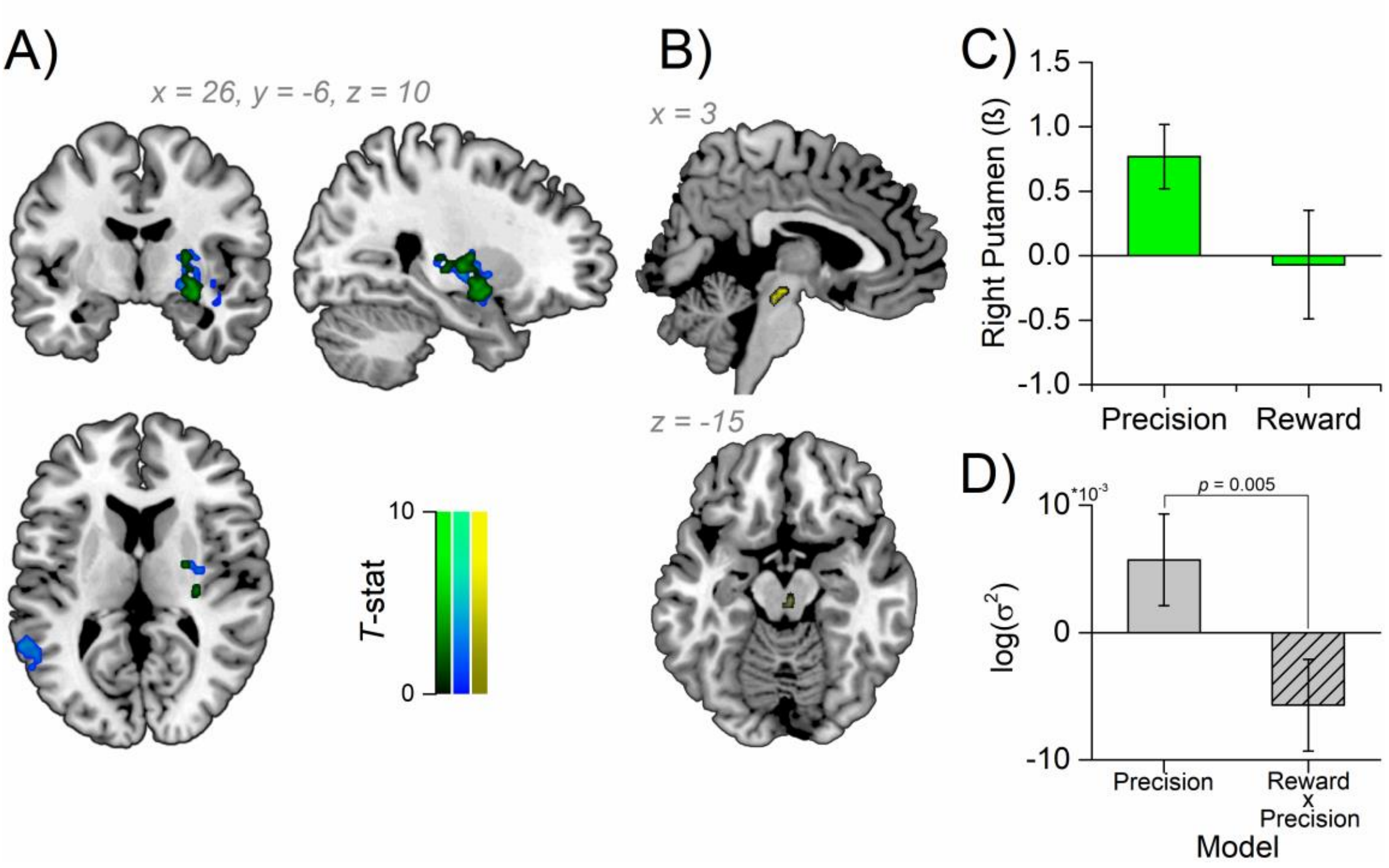

**Figure 3**

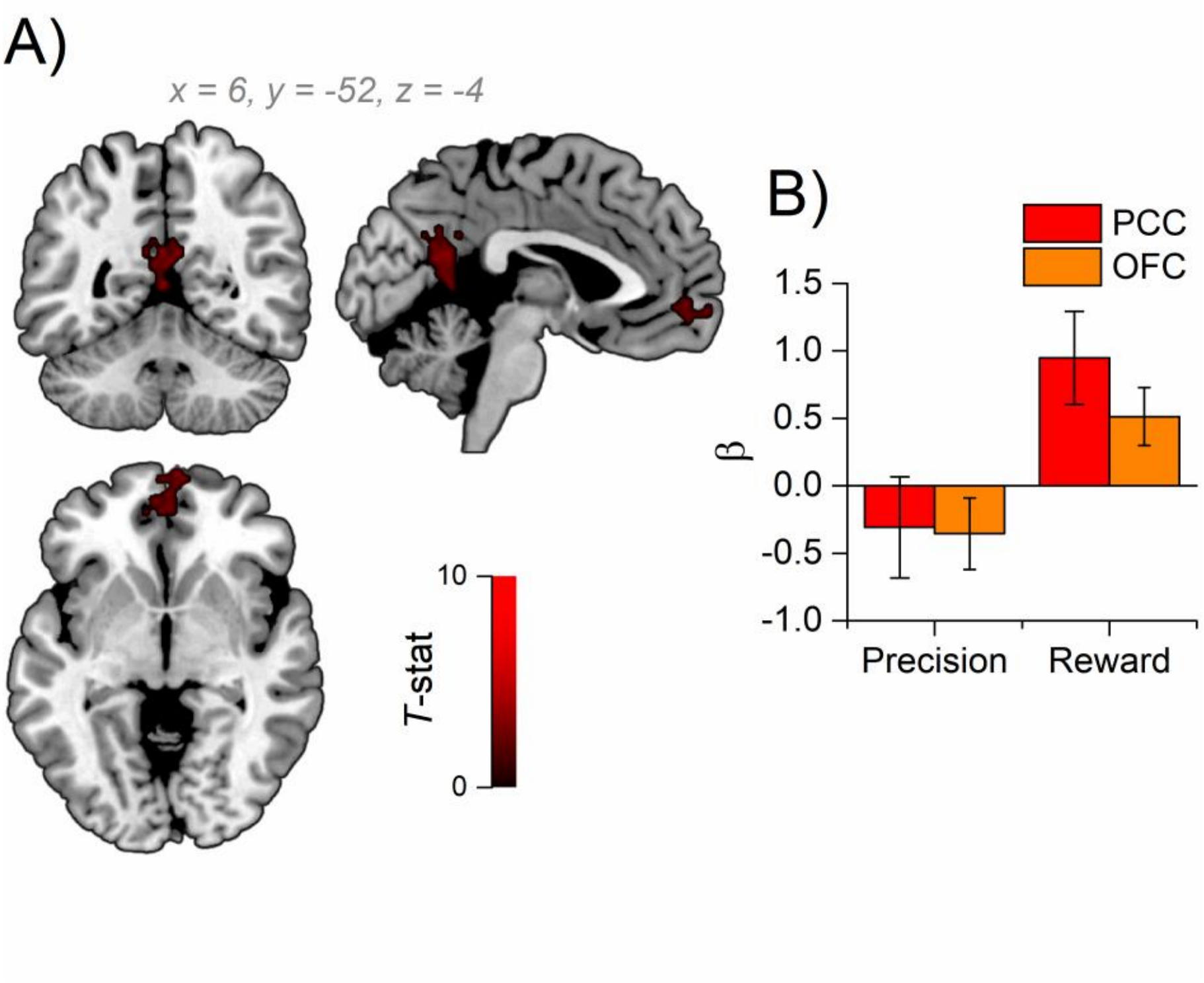

**Figure 4**

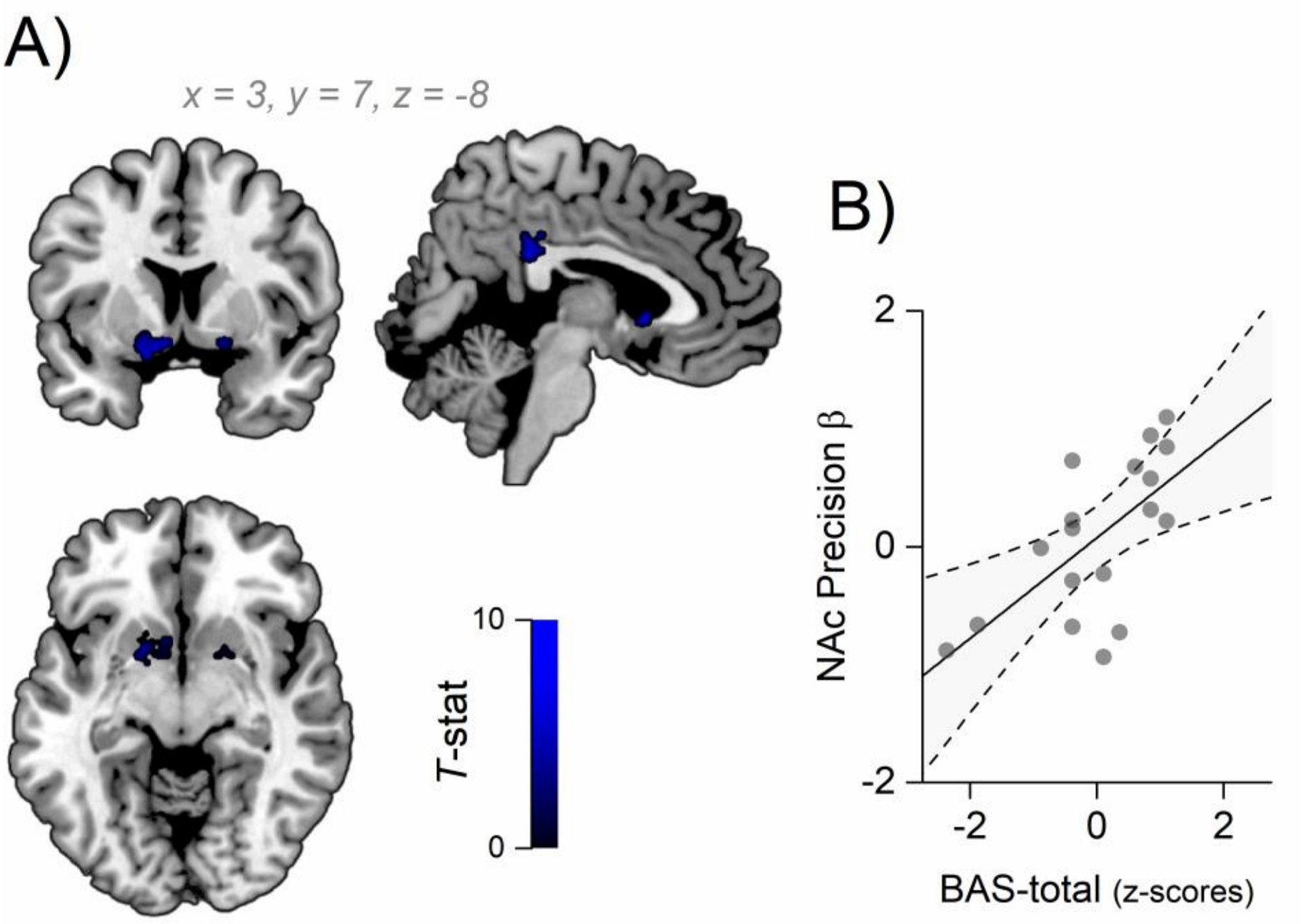